\documentclass[aps, pra, preprint, groupedaddress, amsfonts,
               amsmath, amssymb, nofootinbib]{revtex4-2}
\usepackage{microtype}
\usepackage{graphicx}
\usepackage{epstopdf}
\usepackage[utf8]{inputenc}
\usepackage[T1]{fontenc}
\usepackage[usenames,dvipsnames]{xcolor}
\usepackage[hidelinks]{hyperref}
\usepackage{lmodern}
\usepackage{comment}
\usepackage{subcaption}
\usepackage{float}

\usepackage{color}% \textcolor{#1}{#2}

\begin{document}
\title{Interatomic Coulombic decay initiated by electron removal and excitation processes in helium ion and argon dimer collisions}

\author{Darij Starko}
\email{dmstarko@yorku.ca}
\thanks{Author to whom any correspondence should be addressed.}
\affiliation{Department of Physics and Astronomy, York University, Toronto, Ontario M3J 1P3, Canada}
\author{Tom Kirchner}
\email{tomk@yorku.ca}
\affiliation{Department of Physics and Astronomy, York University, Toronto, Ontario M3J 1P3, Canada}

\date{\today}
\begin{abstract}
	The electron removal and excitation channels in argon dimer target and helium ion projectile collision systems that facilitate interatomic Coulombic decay (ICD) are investigated. We implement an independent-atom and independent-electron model of the collision system with the dimer target fixed at its equilibrium bond length and the He$^{2+}$ and He$^{+}$ ion projectiles travelling parallel to the dimer axis at impact energies ranging from 10 keV/amu to 150 keV/amu. The coupled-channel two-center basis generator method for orbital propagation is used within both a frozen atomic target approximation and a dynamic response framework. Given that ICD is facilitated through electron excitation pathways in argon dimers, a statistical technique called determinantal analysis is employed to investigate these channels. The analysis is further subdivided into models that exclude and include projectile charge changes during the collision. Electron configurations of the form Ar$^{+}$($3p^{-2}nl$) offer a pathway to ICD and are investigated, along with other one- and two-electron removal channels that lead to Ar$^{+}$–Ar$^{+}$ fragmentation. We find that the $3d$ excited state is an overall dominant channel for ICD with other excited states ranging from $4s$-$4f$ also being significant contributors. Our study notes differences between static and dynamical potential models across the projectile impact energy range, though of decreasing significance as the impact energy approaches 150 keV/amu. We also find that a He$^{+}$ projectile offers a strong pathway for ICD as the projectile impact energy decreases.
\end{abstract}

\maketitle
\section{Introduction}
\label{sec:intro}
Since the discovery of interatomic Coulombic decay (ICD) in neon clusters~\cite{Marburger03} and dimers~\cite{Jahnke04}, rare gases have been recognized as a favourable platform for investigating this decay mechanism. Rare gas dimers, such as Ar$_{2}$, are compounds formed by weak van der Waals interactions joining a pair of atoms~\cite{najjari2021}. The reason rare gases are considered an excellent target for testing the ICD mechanism is multifold. Firstly, they are a simple system of the same monoatomic gas in the case of neon or argon dimers, and are relatively easily prepared in a laboratory setting. Second, the gas dimers are loosely bonded, and vulnerable to ionization and excitation processes that facilitate ICD to occur through energy transfer~\cite{Jahnke20}. ICD is an ultrafast decay process in which an outer valence electron in an ionized atom de-excites to fill a vacancy, and the energy released from this transition is transferred to the neighbouring atom, inducing ionization~\cite{Jahnke20}.  

In most of the experimental works discussed in the review paper~\cite{Jahnke20}, photo excitation was used to initiate ICD in the target systems. An alternative approach are ion-cluster collisions, that are particularly relevant pathways for producing showers of low-energy electrons through ICD that damage DNA~\cite{Bouda2000,Alizadeh_2015} with leftover radical species that can react with biomolecules in living tissue~\cite{Vendrell_2010,Pettersson_2013}.

ICD can be experimentally detected from the kinetic energy release spectra of the fragmented ions, ideally recorded in coincidence with the ICD electron. Observing ICD is not easy given multiple other competing fragmentation mechanisms such as radiative charge transfer (RCT) and Coulomb explosion (CE)~\cite{Jahnke20}, resulting in similar signals. RCT involves the removal of two valence electrons from one of the two atoms, followed by radiative relaxation to a Ar$^{+}(3p^{-1})$+Ar$^{+}(3p^{-1})$ state. CE in its simplest case is a process in which two singly charged ions are formed from electron removal from each atom, followed by fragmentation. A reason as to why ICD was first experimentally detected in neon clusters and dimers is because of the simple pathway for ICD in these systems along with the clean resulting spectral signal~\cite{Jahnke04}. 2$s$ electron removal from one neon atom provides an energetically sufficient pathway for ICD which has been observed in several experiments, including in ion-dimer collisions~\cite{Kim13}. Although it is a common and simple test-bed for ICD, it is worth exploring different systems in which this decay originates in more intricate and less clear ways, such as in an argon dimer.

Although Ar is also a noble gas with a filled outer electron shell, removal of an inner valence 3$s$ electron provides insufficient relaxation energy to ionize a neighbouring argon atom's valence electron. An energy of 19.6 eV is needed to remove an electron from the neighbouring argon atom, where 15.8 eV is the ionization energy and 3.8 eV is the minimum shared kinetic energy release (KER) between the argon ion fragments and the ICD electron~\cite{Zhang_2025}. The only relaxation pathways that have the requisite energy are those from the Ar$^{+}(3p^{-2}nl)$ configurations, starting with Ar$^{+}(3p^{-2}3d)$ for which the majority of LS coupled states are energetically open~\cite{Kimura_2013}. 

It was noted in Kim \textit{et al.}~\cite{Kim13} that at impact energies of 150 keV/amu of He$^{2+}$ collisions with the argon dimer target, the KER and electron energy ICD spectral signal was less defined than in the case of the neon dimer target. This is due to a more complex electron configuration requirement for ICD generation in the collision system. In this work we investigate the electron dynamics preceding argon dimer fragmentation by He$^{2+}$ and He$^{+}$ impact at energies ranging from 10 keV/amu to 150 keV/amu. This range of energies offers an additional opportunity to examine the transitions in dominant electron-removal pathways from capture to ionization from low to high impact energies. We present a study on the electron dynamics in these collision systems to shed light on the ICD signature for the argon dimer target.

\section{Theoretical Framework}
\label{sec:TF}

We construct a framework for the collision systems under investigation which is based on those described in~\cite{Kirchner_2021,Starko_2025}. Projectile collisions with each atom in the dimer are treated sequentially under the independent atom model (IAM). In the IAM each atom's interaction with the projectile is described through a time-dependent Hamiltonian within the semiclassical approximation~\cite{Kirchner_2021}. For each atom, an independent electron model (IEM) is applied in which the ion-atom collision is treated as a many electron problem with electron-electron correlations neglected. The system is described by a time-dependent Schr{\"o}dinger equation, and a Hamiltonian (expressed in atomic units) of the form
\begin{equation}
	H(t)=-\dfrac{1}{2}\Delta+\upsilon_{T}\left(r\right)+\upsilon_{P}\left(\vec{r},t\right)
	\label{eq:hamiltonian}
\end{equation}
is used~\cite{Dyuman20}. $\upsilon_{T}\left(r\right)$ is a spherically symmetric effective target potential, which includes the nuclear Coulomb potential. This target potential is obtained from an optimized potential method (OPM) calculation at the level of the exchange-only approximation~\cite{ee93}. The term $\upsilon_{P}\left(\vec{r},t\right)$ is the bare Coulomb potential of the helium nucleus in the case of He$^{2+}$ impact. In the case of a He$^{+}$ projectile, the potential is augmented to include a Coulomb potential and a screening potential of the form
\begin{equation}
	\upsilon_{scr}\left(r_{p}\right)=\int d^{3}r_{p}'\dfrac{\varphi_{1s}^{2}\left(r_{p}'\right)}{\left|\vec{r_{p}}-\vec{r_{p}'}\right|}
	\label{eq:scr_pot}
\end{equation}
where $r_{p}=\left|\vec{r}-\vec{R}\left(t\right)\right|$, $\vec{R}\left(t\right)$ is the classical projectile trajectory and $\varphi_{1s}$ is the (numerical) ground-state Hartree-Fock (HF) orbital of neutral helium. This particular projectile potential is chosen so as to provide a good representation of the binding energy of an electron captured into the 1$s$ orbital ($\varepsilon_{1s}=-0.918$ au). The potential $\upsilon_{P}$ behaves as $-2r^{-1}$  at close distances and decays like $-1/r$ asymptotically.

The static potential considered so far has been referred to as the no response model. It is deemed accurate when the projectile impact velocity is large compared to the average orbital velocity $\bar{v}$ of an outer target electron. For the Ar valence electrons, $\bar{v}$ corresponds to an impact energy of approximately 30 keV/amu. At such impact velocities, the electron density of the target potential does not substantially change over the course of the brief collision. However, at low to intermediate projectile impact energies, time-dependent screening effects play an important role. Factoring this in, a non-zero time-dependent potential is used in what is called the response model. To account for target response during ionization and capture throughout the collision in a global fashion, the approach presented in~\cite{tom00} is used. In it the dynamic target potential is described by a linear combination of ionic ground-state potentials, weighted by time-dependent probabilities to produce a fractional charge during the collision~\cite{tom00}. The idea behind the response model is to adjust the atomic potential in the spirit of time-dependent mean-field theory to reflect the fractional ionic character a target atom acquires when electrons are removed from it. The ionic character is regulated by the time-dependent net electron removal probability. The response effects remain marginal so long as zero- and one-fold electron removal dominate, but they may become significant once higher-order removal processes contribute substantially. The projectile impact energy dependence of the response model will be later discussed in section \ref{sec:Results}.

For these potentials, basis states need to be constructed in accordance with the assumptions made so far in order to represent the time-dependent Schr{\"o}dinger equation that describes the ion-atom collision system~\cite{tcbgm} and to extract transition amplitudes to be used in the analysis and submodels presented in \ref{sec:model}. This is done using the two-center basis generator method (TC-BGM) where both target and projectile centres are used for the construction of basis states. The promising aspect of this method is that the model space spanned by the basis functions is dynamically adapted to the time-dependent problem, thereby minimizing the couplings to the Hilbert space not covered by the basis states~\cite{OJKroneisen_1999}.

In the present work, the basis states for argon consist of the $2s$ to $4f$ target orbitals as well as the $1s$ to $4f$ projectile orbitals and a set of pseudo-states~\cite{Kirchner_2021}. The argon $1s$ orbital is not included given the far higher energy requirement to remove a K-shell electron in comparison to those in the L and M shells. In the case of He$^{+}$ impact, there are $N=N_{T}+1$ bound electrons; $N_{T}$ and 1 on the target and projectile respectively. Transition amplitudes for electron changing states are obtained from solving the $N$ single-electron time-dependent Schrödinger equations corresponding to the single-particle Hamiltonian (\ref{eq:hamiltonian}).

The collision system is set up with the argon dimer fixed at a bond length of $R_e=3.76$ a.u.~\cite{Moseley_1977}, while the projectile travels parallel to the dimer axis at impact parameters $b$ ranging from 0.2 to 10 a.u.. Electron dynamics are studied from the sequential ion-atom collisions between the projectile and the argon atoms. Electron removal in coincidence with excitation is of particular interest as a method of facilitating ICD in the argon dimer. Not all electron excitation configurations are energetically able to facilitate ICD in an argon dimer, and a scaling method to account for this is described later in section \ref{sec:model}. At low impact energies, electron removal is practically equivalent with electron capture by the projectile, but this is not the case at projectile speeds above $\bar{v}$. The transition in dominance between capture and ionization processes is accounted for and described in section \ref{sec:model} of this paper.

\section{Model Details}
\label{sec:model}
We build upon the determinantal analysis of~\cite{Starko_2025}, and apply it to an argon dimer target. The multinomial analysis~\cite{Kirchner_2021,Starko_2025} is not examined here given the preference of the more fundamental determinantal approach. As was previously done~\cite{Starko_2025}, we make the assumption that the projectile interacts with each atom independently in a sequential manner where electron capture from the first atom changes the collision dynamics with the second atom. We note that orientation-dependent information is accessible experimentally via measurement of the momenta of the dimer fragments (see, e.g.,~\cite{Siddiki_2023,Kim2014,Iskandar2014} and references cited therein), i.e., a future experimental study that directly compares with our results for one particular orientation seems feasible in principle.

The determinantal analysis is a statistical method in which determinants of single-particle density matrices are computed and the Pauli exclusion principle is accounted for. The electron removal processes from the dimer include the following categories: one electron removal, one electron removal from each atom in the dimer (two-site), and two electron removal from a single atom (one-site), one electron removal and one electron excitation on a single atom. The previously developed fixed-charge and capture models I and II~\cite{Starko_2025} are applied in order to examine the repercussions on electron dynamics in the collision system depending on projectile charge. 

The determinantal analysis and its submodels are further subdivided by considering them with a frozen target potential (no-response) model and a dynamic response model following the details laid out in~\cite{tom00} and applied in~\cite{Starko_2025}.

The dynamical response approach for each analysis and model is conducted as the no-response model, except that the final-state analyses are carried out at eleven, instead of just one, final distances between the projectile and the target, with results averaged. Due to the dynamically changing potential, projecting the solutions onto inital target eigenstates results in fluctuating transition probabilities~\cite{tom00}. This then requires calculating these probabilities by sampling a set of distances (ranging from $z=30$ a.u. to $50$ a.u. in the present work) and then averaging~\cite{Starko_2025}. 

For the He$^{+}$ projectile, capture of an electron results in a neutral atom for which we assume that its interaction with a subsequent atom cannot trigger electron removal, i.e. we set the elastic probability equal to one in this scenario. This means He$^{+}$ is limited to single electron capture. 

\subsection{Determinantal Analysis}
\label{ssec:DA}
This analysis is similar to~\cite{Starko_2025}, but extended to include electron excitation channels as well as a scaling model to account for the increasing importance of direct ionization with increasing projectile impact energy. 

Transition amplitudes are assumed to be described by inner products of Slater determinants, that by design account for the Pauli exclusion principle~\cite{Starko_2025}. The transition from initial-to-final-state electron configurations results in a probability that is equal to the determinant of a single-particle density matrix:
\begin{equation*}
	P_{f_{1}\ldots f_{N_{T}}} = |\left\langle f_{1}\ldots f_{N_{T}}|i_{1}\ldots i_{N_{T}},t_{f}\right\rangle|^2
\end{equation*}
\begin{equation}
	\begin{aligned}
	& =
	\begin{vmatrix}
		\left\langle f_{1}|i_{1}\right\rangle & \cdots & \left\langle f_{1}|i_{N_{T}}\right\rangle \\
		\vdots        & \ddots & \vdots \\
		\left\langle f_{N_{T}}|i_{1}\right\rangle & \cdots & \left\langle f_{N_{T}}|i_{N_{T}}\right\rangle
	\end{vmatrix}
	\times
		\begin{vmatrix}
		\left\langle i_{1}|f_{1}\right\rangle & \cdots & \left\langle i_{1}|f_{N_{T}}\right\rangle \\
		\vdots        & \ddots & \vdots \\
		\left\langle i_{N_{T}}|f_{1}\right\rangle & \cdots & \left\langle i_{N_{T}}|f_{N_{T}}\right\rangle
	\end{vmatrix} \\
	& = 
	\begin{vmatrix}
		\gamma_{11} & \cdots & \gamma_{1N_{T}} \\
		\vdots        & \ddots & \vdots \\
		\gamma_{N_{T}1} & \cdots & \gamma_{N_{T}N_{T}}
	\end{vmatrix} \equiv det(\gamma).
	\end{aligned}
	\label{eq:exclusive}
\end{equation}
Here $\left|i_{j},t_{f}\right\rangle$ is the time-propagated orbital corresponding to the initial state $\left|i_{j}\right\rangle$, $\left|f_{j}\right\rangle$ is a final state, $\gamma_{jk}$ are the density matrix elements computed from the Slater determinants, and $N_{T}$ is the number of active target electrons ($N_{T}$=16 in the present work). The density matrix elements are sums of products of single-particle transition amplitudes from a range of initial to final states:
\begin{equation}
	\gamma_{jk}(t_{f}) = \left\langle f_{j}|\gamma\left(t_{f}\right)|f_{k}\right\rangle =\sum_{i=1}^{N_{T}}\left\langle f_{j}|i,t_{f}\right\rangle \left\langle i,t_{f}|f_{k}\right\rangle =\sum_{i}^{N_{T}}c_{k}^{{i}^{*}}\left(t_{f}\right)c_{j}^{i}\left(t_{f}\right).
\end{equation}

When we explicitly specify the final states of all electrons in the system, as in equation (\ref{eq:exclusive}), the resulting quantity is referred to as an exclusive probability. This represents a fully resolved, complete measurement. In practice, however, such detailed information is rarely accessible. More commonly, we consider the probability for certain electron transitions without fixing the final states of all remaining electrons. This is known as an inclusive probability~\cite{Ludde_1985}. This probability is a sum of exclusive probabilities of some electron states and can be written as:
\begin{equation}
	P_{f_{1}\ldots f_{q}}=\sum_{f_{q+1}<\ldots<f_{N_{T}}}P_{f_{1}\ldots f_{N_{T}}}.
	\label{eq:inclusive}
\end{equation}
Here $q$ out of $N_{T}$ electrons are in the final state $\left|f_{1}\ldots f_{q}\right\rangle$ while the remaining $N_{T}-q$ electrons are left unspecified~\cite{Starko_2025}. In accordance with equation (\ref{eq:exclusive}), $P_{f_{1}\ldots f_{q}}$ is the determinant of one $q\times q$ matrix corresponding to the sub-configuration $\left|f_{1}\ldots f_{q}\right\rangle$. When electrons are removed, it is also necessary to account for the vacancies they leave behind—the hole states. Following (\ref{eq:inclusive}), we formulate an inclusive particle-hole probability as follows:
\begin{equation}
	\begin{aligned}
		\begin{array}{c}
			P_{f_{1}\ldots f_{q}}^{\bar{f_{1}}\ldots\bar{f_{k}}}\equiv\sum_{f_{q+1}<\ldots<f_{N_{T}}}^{\bar{f_{1}}\ldots \bar{f_{k}}}P_{f_{1}\ldots f_{N}}\\
			=P_{f_{1}\ldots f_{q}}-{\displaystyle \sum_{l=1}^{k}}P_{f_{1}\ldots f_{q}\bar{f_{l}}}+{\displaystyle \sum_{l_{1}<l_{2}}^{k}}P_{f_{1}\ldots f_{q}\bar{f_{l_{1}}}\bar{f_{l_{2}}}}-\cdots,
		\end{array}
	\end{aligned}
	\label{eq:holes}
\end{equation}
where $\bar{f_{i}}$ are the hole states, and the resulting particle–hole probability appears as an alternating series of positive and negative contributions~\cite{Ludde_1985,AToepfer_1985}. This framework enables the calculation of the electron‑removal channels relevant to this work, as discussed in the sections that follow.

\subsubsection{Removal without Electron Excitation}
  
The inclusive model can be extended to include the projectile ion in the analysis. For He$^{2+}$ impact, we consider two models that take into account the capture of an electron by the projectile to form a He$^{+}$ ion and its interaction with the second atom in the dimer; models I and II~\cite{Starko_2025}.

Model I incorporates the above inclusive probability formulation, yet using transition amplitudes computed from interactions between He$^{+}$ and an Ar atom in addition to those for He$^{2+}-$Ar collisions. The channel probabilities are computed in an analogous manner as in the fixed-charge model approach. 

Model II has fundamentally the same structure as model I when computing final probabilities, however the density matrix elements include initial states from the projectile ion. We only consider the $1s$ state, due to that being the most relevant state an electron will occupy in the ion after capture. We can account for this by writing
\begin{equation}
	\gamma_{kj}=\sum_{i=1}^{N_{T}+1}\left\langle f_{k}|i,t_{f}\right\rangle \left\langle i,t_{f}|f_{j}\right\rangle.
\end{equation}

The inclusion of the transition from the projectile to the second target atom is indicated by the +1 above the sum.

Applying the same techniques as were done for the neon dimer in~\cite{Starko_2025}, all two-site electron removal probabilities can be written as:

\begin{equation}
\begin{array}{c}
	P_{s^{-1},s^{-1}}^{\rm dimer}=P_{s^{-1}}P_{s^{-1}}\\
	P_{p^{-1},p^{-1}}^{\rm dimer}=P_{p^{-1}}P_{p^{-1}}\\
	P_{s^{-1},p^{-1}}^{\rm dimer}=2P_{s^{-1}}P_{p^{-1}}\\
	P_{s^{-2}}^{\rm dimer}=2P_{s^{-2}}P_{\rm elastic}\\
	P_{p^{-2}}^{\rm dimer}=2P_{p^{-2}}P_{\rm elastic}\\
	P_{s^{-1}p^{-1}}^{\rm dimer}=2P_{s^{-1}p^{-1}}P_{\rm elastic}.
\end{array}
\label{eq:int_p}
\end{equation}

The subscripts in (\ref{eq:int_p}) have the principle quantum number $n=3$ dropped for convenience. The terms $s^{-1},s^{-1}$ and $s^{-2}$ denote two-site and one-site $3s$ electron removal respectively, with the comma indicating a two-site process. The same notation applies to the remaining electron‑removal channels listed in (\ref{eq:int_p}).

This is repeated for capture model I, where $*$ indicates a He$^{+}$ projectile interaction probability
\begin{equation}
\begin{array}{c}
	P_{s^{-1},s^{-1}}^{\rm dimer}=P_{s^{-1}}P_{s^{-1}}^{*}\\
	P_{p^{-1},p^{-1}}^{\rm dimer}=P_{p^{-1}}P_{p^{-1}}^{*}\\
	P_{s^{-1},p^{-1}}^{\rm dimer}=P_{s^{-1}}P_{p^{-1}}^{*}+P_{p^{-1}}P_{s^{-1}}^{*}\\
	P_{s^{-2}}^{\rm dimer}=P_{s^{-2}}+P_{\rm elastic}P_{s^{-2}}\\
	P_{p^{-2}}^{\rm dimer}=P_{p^{-2}}+P_{\rm elastic}P_{p^{-2}}\\
	P_{s^{-1}p^{-1}}^{\rm dimer}=P_{s^{-1}p^{-1}}+P_{\rm elastic}P_{s^{-1}p^{-1}}.
	\label{eq:cap_m0_p}
\end{array}
\end{equation}

The last three probabilities in (\ref{eq:cap_m0_p}) assume that interactions between a neutral He atom and Ar will be negligible.

For a He$^{+}$ projectile, only single electron removal processes are allowed in the capture model for the reasons stated earlier. 

\subsubsection{Electron Excitation}
In the argon dimer target, electron excitation plays a significantly more prominent role for facilitating ICD in the energy regime considered here than in the case of the neon dimer target~\cite{Kim13}. These electron excitations occur through single–electron removal accompanied by the promotion of another electron to a higher orbital, leading to valence–shell configurations of the form $3p^{-2}nl$.
The probability of populating a specific excited state $nl$ is evaluated using the inclusive formalism introduced in the previous section, equation (\ref{eq:holes}) in particular. As an illustrative example, the probability for one electron to be removed from the $3p_{0}$ subshell while another electron is excited to an $nl$ orbital is given by
\begin{equation}
	P_{3s^{2},3p_{1g}^{2},3p_{1u}^{2},nl\uparrow}^{3\bar{p_{0}}^{2}}
	= P_{3s^{2},3p_{1g}^{2},3p_{1u}^{2},nl\uparrow}
	- P_{3s^{2},3p_{0}\downarrow,3p_{1g}^{2},3p_{1u}^{2},nl\uparrow},
	\label{eq:arg_exc_inc}
\end{equation}

where $nl\uparrow$ denotes the excited state with spin up (chosen without loss of generality). The first term on the right-hand side in (\ref{eq:arg_exc_inc}) represents configurations containing two $3p$ vacancies together with one electron in the excited orbital. The second term accounts for configurations with a single spin–up vacancy and an electron promoted to $nl$. All such combinations contributing to a given excited state $nl$ are evaluated explicitly. 

The combination of states associated with a particular excited configuration is determined from its term symbol ${}^{2S+1}L$, where $L$ and $S$ are the orbital and spin angular momentum quantum numbers respectively, and $\left(2S+1\right)$ is the spin multiplicity. The  number of states for a particular term symbol is $\left(2S+1\right)\left(2L+1\right)$, with \{$L|S=0,P=1,D=2,F=3,\ldots$\} and \{$S=0,1/2,1,3/2,..$\}~\cite{NIST_ASD}.

Not all LS coupled states for a particular configuration exceed the energetic threshold to initiate ICD, and this needs to be be accounted for. A procedure for scaling the probability given an energy threshold cutoff is developed.

The procedure is implemented by first summing all configurations that contain two $3p$ vacancies and one excited electron, without imposing the impossibility of spin flips. The calculation is then repeated for configurations with only a single $3p$ vacancy and the excited electron. This ensures that all relevant electron configurations are included. After these contributions are obtained, each total is scaled by the fraction of states whose excitation energies exceed the minimum threshold required for ICD. For instance, for $3d$ excited state the corresponding scaling factor is $42/90$. To calculate the scaling factor it is necessary to first determine the total number of states of Ar$^{+}(3p^{-2}3d)$ type, which is 150. However, 60 of these correspond to spin quartet states which are not accessible without spin flips and have to be removed from the list of allowed states, leaving a total of 90. Out of these 90 states, only 42 exceed the necessary energy threshold for facilitating ICD~\cite{NIST_ASD}.

In the case of the $n=4$ excited states, the scaling factors are 1/9 for $4s$ and 1 for the remaining excitation channels, i.e. all LS coupled states lie above the ICD threshold for the latter.

Once this scaling is obtained, the probability for electron removal with simultaneous electron excitation $P_{p^{-2}nl}$ is used to supplement (\ref{eq:int_p}), (\ref{eq:cap_m0_p}) and (\ref{eq:hecap_m0_p}) with the following dimer probabilities for the fixed-charge model:

\begin{equation}
	P_{p^{-2}nl}^{\rm dimer}=2P_{p^{-2}nl}P_{\rm elastic},
	\label{eq:he2fcm_exc}
\end{equation}
for the capture models with the He$^{2+}$ projectile:
\begin{equation}
	P_{p^{-2}nl}^{\rm dimer}=P_{p^{-2}nl}P_{\rm elastic}^{*}+P_{\rm elastic}P_{p^{-2}nl}
	\label{eq:he2cap_exc}
\end{equation}
and for the capture models with the He$^{+}$ projectile
\begin{equation}
		P_{p^{-2}nl}^{\rm dimer}=P_{p^{-2}nl}+P_{\rm elastic}P_{p^{-2}nl}.
		\label{eq:hecap_exc}
\end{equation}

\subsubsection{Energy Scaling}

The calculated probabilities obtained from the capture models for both He$^{2+}$ and He$^{+}$ projectiles have thus far relied on the assumption that for projectile impact energies less than $\bar{v}$, electron capture strongly dominates over ionization. As the projectile impact energy increases into the regime greater than $\bar{v}$, this assumption ceases to hold, and ionization must be explicitly incorporated. 

A simple method of factoring in the balance between electron ionization and capture processes is to incorporate their relative contributions to the total electron removal in the model. These ratios are obtained from

\begin{align}
	p_{net}^{rem}&=N_{T}-\sum_{i=1}^{N_{T}}\sum_{j=1}^{J_{T}}|c_{j}^{i}|^{2}, &
	p_{net}^{cap}&=\sum_{i=1}^{N_{T}}\sum_{j=1}^{J_{P}}|c_{j}^{i}|^{2}, &
	p_{net}^{ion}&=p_{net}^{rem}-p_{net}^{cap},
	\label{eq:netremprobs}
\end{align}

where $J_{T}$ and $J_{P}$ are the number of target and projectile basis states respectively, and $p_{net}^{rem}$, $p_{net}^{cap}$, and $p_{net}^{ion}$ are the net electron removal, capture and ionization probabilities respectively. Given (\ref{eq:netremprobs}), a balance between ionization and capture probabilities is found utilizing the fixed-charge and capture model probabilities from (\ref{eq:int_p}),(\ref{eq:cap_m0_p}) and (\ref{eq:hecap_m0_p}). The fixed-charge model assumes no change in projectile charge, effectively representing an ionization model, while capture models I and II describe a capture process. Scaling the probabilities between these two types of models is done through the use of the ratio of probabilities in (\ref{eq:netremprobs}) to give a final electron probability $P$ for a particular projectile impact energy as
\begin{equation}
	P=\alpha P_{FCM}+\beta P_{CM}
	\label{eq:ratio_prob}
\end{equation}

where $P_{FCM}$, $P_{CM}$ are fixed-charge and capture model probabilities respectively, while $\alpha$ and $\beta$ are the ratio of ionization and capture with respect to removal, defined as
\begin{align}
	\alpha &= \dfrac{p_{net}^{ion}}{p_{net}^{rem}}, &
	\beta &= \dfrac{p_{net}^{cap}}{p_{net}^{rem}}.
\end{align}

At low projectile impact energies, capture dominates the removal channel, and the contribution from $P_{CM}$ is large, while $P_{FCM}$ provides only a minor correction. As the impact energy increases, the situation reverses: ionization becomes the dominant mechanism, and the relative weight shifts toward the fixed-charge model. Going forward, the capture models will be reclassified as the modified capture models I (MCMI) and II (MCMII) to account for this weighting scheme that balances ionization and capture dominance.

\section{Results and discussion}
\label{sec:Results}
% Points
We first consider the He$^{2+}$ projectile, then He$^{+}$, and we focus on the $3p^{-2}nl$ channels, which are associated with ICD. Going forward, in all figures references to electron channels will not include the principal quantum number $n=3$ in front (i.e. we replace $3s$ with $s$ etc.); res will refer to the response model and hep to the He$^{+}$ projectile being used in the analysis. The single electron removal and electron excitation channels will be labeled $pp3d$ for example, while the sum of all excitations is referred to as $ppnl$. To simplify references with respect to analyses, we refer to the determinal analysis as DA. Models for an analysis also use acronyms (e.g. fixed-charge model is FCM; modified capture model I is MCMI). Identifying a particular analysis and model uses a XX.YYY format, where XX refers to the analysis and YYY the model (e.g. DA.FCM stands for the fixed-charge model in the determinantal analysis)~\cite{Starko_2025}.

\subsection{He$^{2+}$ Projectiles}
We start by looking at the probability distributions at projectile impact energies of 10 keV/amu and 150 keV/amu across a range of impact parameters.

\begin{figure}[H]
	\begin{subfigure}{.48\textwidth}
		\centering
		\includegraphics[width=1\linewidth]{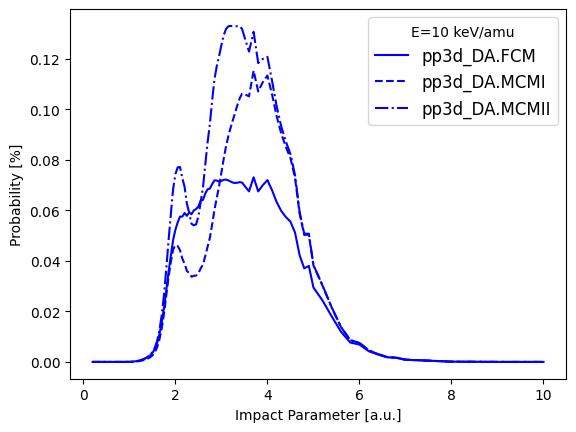}
		\caption{10 keV/amu}
	\end{subfigure}
	\begin{subfigure}{.48\textwidth}
		\centering
		\includegraphics[width=1\linewidth]{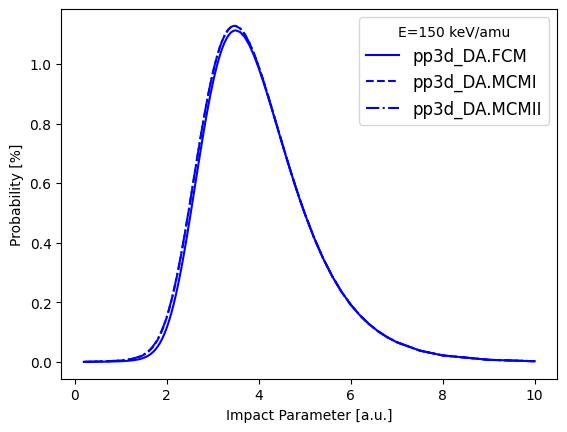}
		\caption{150 keV/amu}
	\end{subfigure}
	\caption{He$^{2+}$-Ar$_{2}$ ion-dimer collisions in parallel orientation at a) 10 keV/amu and b) 150 keV/amu; $3p$ electron removal and $3p$ electron excitation to $3d$ state  probability comparing DA.FCM (solid line) vs DA.MCMI (dashed line) and DA.MCMII (dotted-dashed line) in the no-response model.}
	\label{fig:modelcomp}
\end{figure}

Figure \ref{fig:modelcomp} shows for the $3p^{-2}3d$ channel that the models converge at 150 keV/amu, while displaying distinct differences at the lowest energy (10 keV/amu). The high-energy convergence behaviour is expected given the dominance of electron ionization over capture at higher energies, as a consequence of which the modified capture model is effectively the fixed-charge model. It is noteworthy that at 10 keV/amu, all the models have very small probabilities, resulting in less smooth curves in comparison to those at 150 keV/amu. 

\begin{figure}[H]
	\begin{subfigure}{.48\textwidth}
		\centering
		\includegraphics[width=1\linewidth]{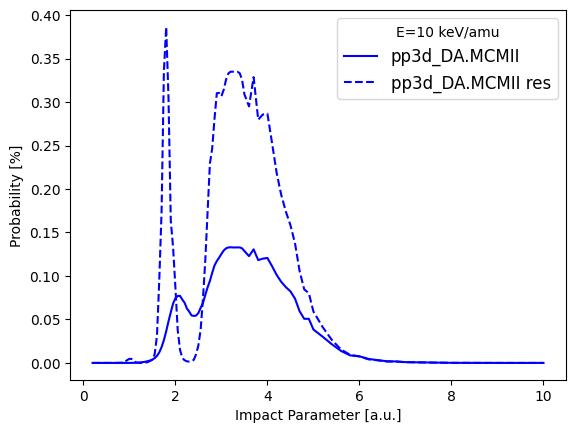}
		\caption{10 keV/amu}
	\end{subfigure}
	\begin{subfigure}{.48\textwidth}
		\centering
		\includegraphics[width=1\linewidth]{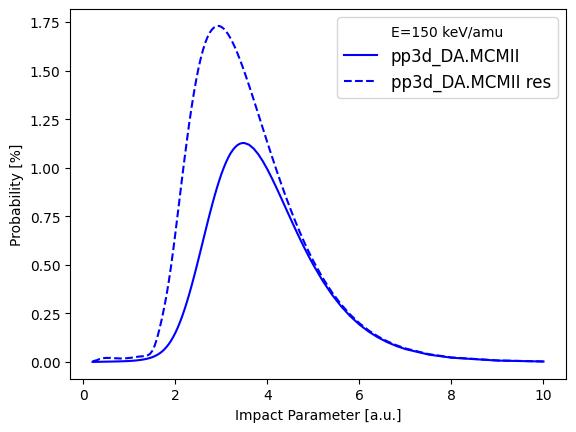}
		\caption{150 keV/amu}
	\end{subfigure}
	\caption{He$^{2+}$-Ar$_{2}$ ion-dimer collisions in parallel orientation at a) 10 keV/amu and b) 150 keV/amu; $3p$ electron removal and $3p$ electron excitation to $3d$ state probability comparison of DA.MCMII between no-response (solid line) and response (res) (dashed line) models.}
	\label{fig:rescomp}
\end{figure}

In Figure \ref{fig:rescomp} for the same channel, differences in the probability distributions due to the use of a time-dependent potential are observed at both 10 keV/amu and 150 keV/amu.

Although there are differences between no-response and response models across the impact energy range, it is notable that the models are more similar at 150 keV/amu. This is an expected behaviour as the dynamic potential evolves more when the target has a longer exposure to the projectile, and as impact energy increases the projectile passes by the target much quicker leaving less impact. If we were to explore this system at impact energies larger than 150 keV/amu, we would expect convergence between response and no-response.

\begin{figure}[H]
	\begin{subfigure}{.48\textwidth}
		\centering
		\includegraphics[width=1\linewidth]{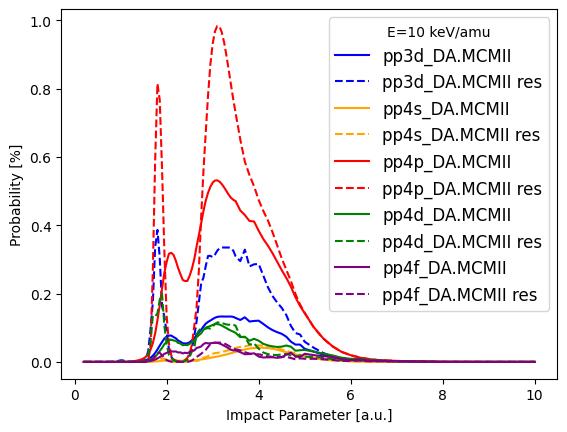}
		\caption{10 keV/amu}
	\end{subfigure}
	\begin{subfigure}{.48\textwidth}
		\centering
		\includegraphics[width=1\linewidth]{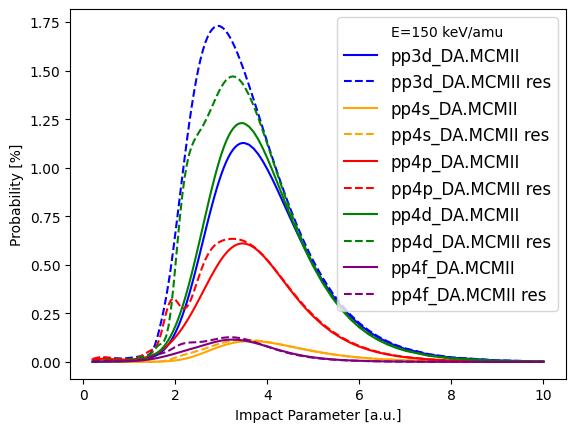}
		\caption{150 keV/amu}
	\end{subfigure}
	\caption{He$^{2+}$-Ar$_{2}$ ion-dimer collisions in parallel orientation at a) 10 keV/amu and b) 150 keV/amu; $3p$ electron removal and $3p$ electron excitation to $3d,4s,4p,4d,$ and $4f$ state probability comparisons of DA.MCMII between no-response (solid line) and response (res) (dashed line) models.}
	\label{fig:excited}
\end{figure}

In Figure \ref{fig:excited} we show results for all excitation channels ($3d$ to $4f$) considered. It can be seen that various channels dominate at different energies. At 10 keV/amu impact energy, the $4p$ excitation channel is dominant in both no-response and response models while $3d$ is the second largest. However, at 150 keV/amu impact energy the $3d$ and $4d$ excitation channels dominate, with $3d$ response model being the most dominant. We note that there is a clear difference in structure between both sets of plots in a) and b), where similarly as in Figure \ref{fig:rescomp} the excitation channels all exhibit similar structures and main peaks near the same impact parameter $b$ in the range of $3-3.5$ a.u. for both response and no-response models. We also see the more fractured probability curves at 10 keV/amu where each channel differentiates itself from the others.

\begin{figure}[H]
	\begin{subfigure}{.48\textwidth}
		\centering
		\includegraphics[width=1\linewidth]{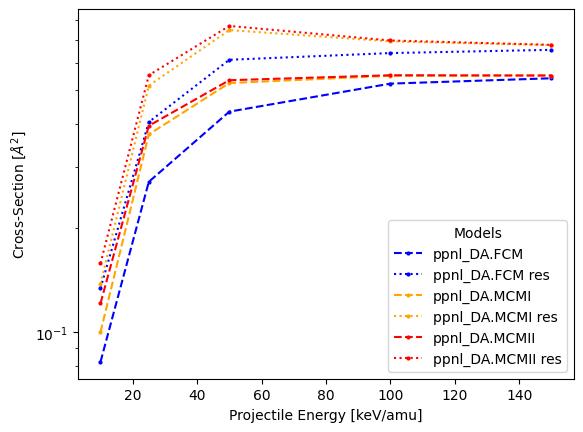}
		\caption{Sum of $3d$ to $4f$ $3p^{-2}nl$ cross-sections}
	\end{subfigure}
	\begin{subfigure}{.48\textwidth}
		\centering
		\includegraphics[width=1\linewidth]{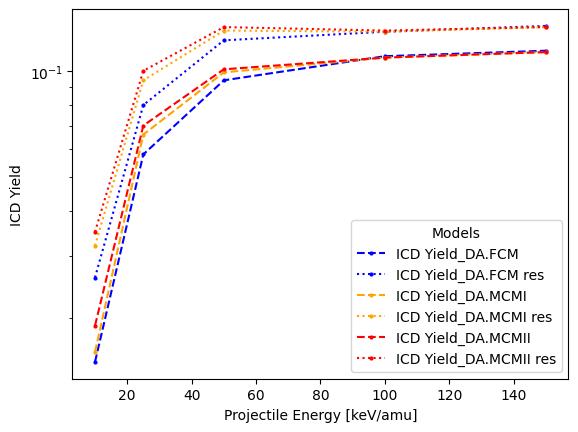}
		\caption{ICD Yield}
	\end{subfigure}
	\caption{He$^{2+}$-Ar$_{2}$ ion-dimer collisions in parallel orientation from 10 keV/amu to 150 keV/amu; a) $3p$ electron removal and $3p$ electron excitation to $nl$ state cross-sections comparison of DA.FCM, DA.MCMI and DA.MCMII between no-response (solid line) and response (res) (dashed line) models b) ICD yield in Ar$^{+}$-Ar$^{+}$ dimer fragmentation comparison of DA.FCM, DA.MCMI and DA.MCMII models between no-response (solid line) and response (res) (dashed line) models}
	\label{fig:he2penergies}
\end{figure}

Figure \ref{fig:he2penergies} shows the sum electron excitation cross-section channels and ICD yield over the range of 10 keV/amu to 150 keV/amu. We define the ICD yield as the fraction of resulting Ar$^{+}$-Ar$^{+}$ fragmentation processes that originate from ICD facilitating channels, compared to all channel cross-sections that result in this fragmentation, including all two-site two-electron removals that result in CE and RCT from $3p^{-2}$ removal. As can be seen, the ICD yield shows a similar pattern as the total cross-section across the projectile impact energy range, plateauing towards 150 keV/amu with capture and fixed-charge models converging as excpected. We also notice a gap in cross-sections and ICD yield between the response and no-response models, from low to high impact energies. This gap appears to slowly narrow as impact energy increases, and is expected to eventually vanish, as faster projectile speeds reduce the impact of dynamical response.

Results from~\cite{Kim13} at 150 keV/amu for He$^{2+}$ projectiles indirectly indicated a high ICD yield due to a large ICD to continuum electron ratio. It does appear that our results are lower than theirs, though this may be due to our results being inclusive in the final projectile charge state, while the data of~\cite{Kim13} were recorded in coincidence with He$^{+}$ production. In addition, our focus on parallel ion-dimer collisions inhibits a quantitative comparison.

\subsection{He$^{+}$ Projectiles}

\begin{figure}[H]
	\begin{subfigure}{.48\textwidth}
		\centering
		\includegraphics[width=1\linewidth]{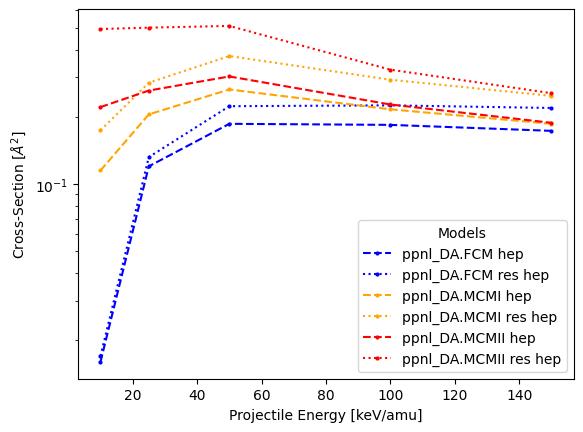}
		\caption{Sum of $3d$ to $4f$ $3p^{-2}nl$ cross-sections}
	\end{subfigure}
	\begin{subfigure}{.48\textwidth}
		\centering
		\includegraphics[width=1\linewidth]{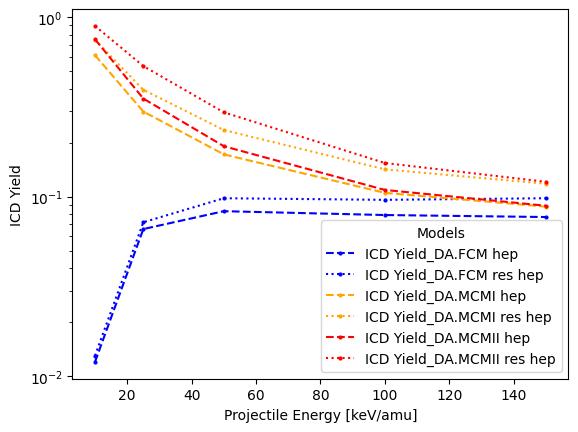}
		\caption{ICD Yield}
	\end{subfigure}
	\caption{He$^{+}$-Ar$_{2}$ ion-dimer collisions in parallel orientation from 10 keV/amu to 150 keV/amu; a) $3p$ electron removal and $3p$ electron excitation to $nl$ state cross-sections comparison of DA.FCM, DA.MCMI and DA.MCMII between no-response (solid line) and response (res) (dashed line) models b) ICD yield in Ar$^{+}$-Ar$^{+}$ dimer fragmentation comparison of DA.FCM, DA.MCMI and DA.MCMII models between no-response (solid line) and response (res) (dashed line) models}.
	\label{fig:hepenergies}
\end{figure}

We repeat these tests with a He$^{+}$ projectile for the same impact energies and apply the same models. The cross-sections and ICD yield ranging from 10 keV/amu to 150 keV/amu are displayed in Figure \ref{fig:hepenergies}.

We see a substantially different behaviour of cross-section and yield. The cross-sections for the capture models exhibit similar structures across models with response results being larger than their no-response counterparts across the entire projectile impact energy range. More notably, the ICD yield for the capture models approaches unity at 10 keV/amu projectile energy, and decays toward high energy. This is very different from the results shown in Figure \ref{fig:he2penergies} for He$^{2+}$ projectiles. The reason is that He$^{+}$ is limited to single electron capture in CMI and CMII (i.e.,Ar$^{+}$-Ar$^{+}$ fragmentation via CE or RCT is impossible).  

\section{Concluding remarks}
\label{sec:conclusions}

Electron removal with electron excitation processes in helium ion collisions with argon dimers were studied with an emphasis on channels that facilitate ICD. In particular, electron removal channels that result in Ar$^{+}$-Ar$^{+}$ fragmentation were studied as this fragmentation pathway is the possible byproduct of the ICD mechanism. We expanded upon the work done in~\cite{Starko_2025} by incorporating in the determinantal analysis electron excitation as well as balancing capture and ionization dominance based on projectile impact energy. The  collisions and electron dynamics were examined over a range of projectile impact energies from 10 keV/amu to 150 keV/amu. We found that as the energy decreases, the differences in models increases while the time-dependent potential model retains dissimilarities across the range of energies examined with respect to the time-independent potential model. We also observed that a He$^{+}$ projectile capturing a target $3p$ electron and simultaneously exciting another $3p$ electron into a $nl$ state results in a near pure ICD yield in the Ar$^{+}$-Ar$^{+}$ fragmentation channel at 10 keV/amu impact energy, when considered in the modified capture model. These results provide opportunities for experiments to expand upon their investigations in the future.

This theoretical work can be further expanded upon in the future by studying orientation dependence of electron removal by changing the projectile path direction with respect to the dimer axis. Studies can also focus on more complex targets such as water molecule dimers as a more practical example of ICD emerging in biological systems.

\section*{Data Availability}

The datasets generated during and/or analysed during the current study
are available from the corresponding author upon reasonable request.

\newpage
\begin{acknowledgments}
Financial support from the Natural Sciences and Engineering Research Council of Canada (NSERC) 
(RGPIN-2025-06277) is gratefully acknowledged. 
\end{acknowledgments}

%
% BibTeX users please use
%\bibliographystyle{plain}
%\bibliographystyle{unsrt}

\bibliography{icd}

@Article{Bouda2000,
  author  = {Bouda\"iffa, Badia and Cloutier, Pierre and Hunting, Darel and Huels, Michael A. and Sanche, Léon},
  title   = {Resonant Formation of {DNA} Strand Breaks by Low-Energy (3 to 20 {eV}) Electrons},
  journal = {Science},
  year    = {2000},
  volume  = {287},
  number  = {5458},
  pages   = {1658-1660},
  url     = {http://www.sciencemag.org/content/287/5458/1658.abstract},
}

@Article{Marburger03,
  author    = {Marburger, S. and Kugeler, O. and Hergenhahn, U. and M\"oller, T.},
  title     = {{Experimental Evidence for Interatomic Coulombic Decay in Ne Clusters}},
  journal   = {Phys. Rev. Lett.},
  year      = {2003},
  volume    = {90},
  pages     = {203401},
  month     = {May},
  doi       = {10.1103/PhysRevLett.90.203401},
  issue     = {20},
  numpages  = {4},
  publisher = {American Physical Society},
  url       = {https://link.aps.org/doi/10.1103/PhysRevLett.90.203401},
}

@article{Jahnke04,
  title = {Experimental Observation of Interatomic Coulombic Decay in Neon Dimers},
  author = {Jahnke, T. and Czasch, A. and Sch\"offler, M. S. and Sch\"ossler, S. and Knapp, A. and K\"asz, M. and Titze, J. and Wimmer, C. and Kreidi, K. and Grisenti, R. E. and Staudte, A. and Jagutzki, O. and Hergenhahn, U. and Schmidt-B\"ocking, H. and D\"orner, R.},
  journal = {Phys. Rev. Lett.},
  volume = {93},
  issue = {16},
  pages = {163401},
  numpages = {4},
  year = {2004},
  month = {Oct},
  publisher = {American Physical Society},
  doi = {10.1103/PhysRevLett.93.163401},
  url = {https://link.aps.org/doi/10.1103/PhysRevLett.93.163401}
}

@Article{Jahnke20,
  author  = {Jahnke, Till and Hergenhahn, Uwe and Winter, Bernd and Dörner, Reinhard and Frühling, Ulrike and Demekhin, Philipp V. and Gokhberg, Kirill and Cederbaum, Lorenz S. and Ehresmann, Arno and Knie, André and Dreuw, Andreas},
  title   = {{Interatomic and Intermolecular Coulombic Decay}},
  journal = {Chemical Reviews},
  year    = {2020},
  volume  = {120},
  number  = {20},
  pages   = {11295-11369},
  url     = {https://doi.org/10.1021/acs.chemrev.0c00106},
}

@Article{Kim13,
  author    = {Kim, H.-K. and Gassert, H. and Sch\"offler, M. S. and Titze, J. N. and Waitz, M. and Voigtsberger, J. and Trinter, F. and Becht, J. and Kalinin, A. and Neumann, N. and Zhou, C. and Schmidt, L. Ph. H. and Jagutzki, O. and Czasch, A. and Merabet, H. and Schmidt-B\"ocking, H. and Jahnke, T. and Cassimi, A. and D\"orner, R.},
  title     = {{Ion-impact-induced interatomic Coulombic decay in neon and argon dimers}},
  journal   = {Phys. Rev. A},
  year      = {2013},
  volume    = {88},
  pages     = {042707},
  month     = {Oct},
  doi       = {10.1103/PhysRevA.88.042707},
  issue     = {4},
  numpages  = {9},
  publisher = {American Physical Society},
  url       = {https://link.aps.org/doi/10.1103/PhysRevA.88.042707},
}

@Article{Dyuman20,
  author    = {Bhattacharya, Dyuman and Kirchner, Tom},
  title     = {{Strengthening the case for interatomic Coulomb decay as a subdominant reaction channel in slow ${\mathrm{O}}^{3+}\text{\ensuremath{-}}{\mathrm{Ne}}_{2}$ collisions with independent-atom-model coupled-channel calculations}},
  journal   = {Phys. Rev. A},
  year      = {2020},
  volume    = {102},
  pages     = {062816},
  month     = {Dec},
  doi       = {10.1103/PhysRevA.102.062816},
  issue     = {6},
  numpages  = {9},
  publisher = {American Physical Society},
  url       = {https://link.aps.org/doi/10.1103/PhysRevA.102.062816},
}

@article{tom00,
  title = {Time-dependent screening effects in ion-atom collisions with many active electrons},
  author = {Kirchner, T. and Horbatsch, M. and L\"udde, H. J. and Dreizler, R. M.},
  journal = {Phys. Rev. A},
  volume = {62},
  issue = {4},
  pages = {042704},
  numpages = {13},
  year = {2000},
  month = {Sep},
  publisher = {American Physical Society},
  doi = {10.1103/PhysRevA.62.042704},
  url = {https://link.aps.org/doi/10.1103/PhysRevA.62.042704}
}

@Article{tcbgm,
  author  = {M. Zapukhlyak and T. Kirchner and H. J. L\"udde and S. Knoop and R. Morgenstern and R. Hoekstra},
  title   = {Inner- and outer-shell electron dynamics in proton collisions with sodium atoms},
  journal = {J. Phys. B},
  year    = {2005},
  volume  = {38},
  number  = {14},
  pages   = {2353},
  doi     = {10.1088/0953-4075/38/14/003},
  url     = {https://doi.org/10.1088%2F0953-4075%2F38%2F14%2F003},
}

@article{ee93,
  title = {Accurate optimized-potential-model solutions for spherical spin-polarized atoms: Evidence for limitations of the exchange-only local spin-density and generalized-gradient approximations},
  author = {Engel, E. and Vosko, S. H.},
  journal = {Phys. Rev. A},
  volume = {47},
  issue = {4},
  pages = {2800--2811},
  numpages = {0},
  year = {1993},
  month = {Apr},
  publisher = {American Physical Society},
  doi = {10.1103/PhysRevA.47.2800},
  url = {https://link.aps.org/doi/10.1103/PhysRevA.47.2800}
}

@Article{najjari2021,
  author    = {Najjari, B. and Wang, Z. and Voitkiv, A. B.},
  title     = {Probing the Helium Dimer by Relativistic Highly Charged Projectiles},
  journal   = {Phys. Rev. Lett.},
  year      = {2021},
  volume    = {127},
  pages     = {203401},
  month     = {Nov},
  doi       = {10.1103/PhysRevLett.127.203401},
  issue     = {20},
  numpages  = {5},
  publisher = {American Physical Society},
  url       = {https://link.aps.org/doi/10.1103/PhysRevLett.127.203401},
}

@Article{Kirchner_2021,
  author    = {Kirchner, Tom},
  title     = {{Indication of strong interatomic Coulombic decay in slow ${\mathrm{He}}^{2+}\text{\ensuremath{-}}{\mathrm{Ne}}_{2}$ collisions}},
  journal   = {Journal of Physics B: Atomic, Molecular and Optical Physics},
  year      = {2021},
  volume    = {54},
  number    = {20},
  pages     = {205201},
  month     = oct,
  doi       = {10.1088/1361-6455/ac34e0},
  issn      = {1361-6455},
  publisher = {IOP Publishing},
  url       = {http://dx.doi.org/10.1088/1361-6455/ac34e0},
}

@Article{Siddiki_2023,
  author  = {Siddiki, Md and Tribedi, Lokesh and Misra, Deepankar},
  title   = {Probing the Fragmentation Pathways of an Argon Dimer in Slow Ion–Dimer Collisions},
  journal = {Atoms},
  year    = {2023},
  volume  = {11},
  pages   = {34},
  month   = {02},
  doi     = {10.3390/atoms11020034},
}

@Article{AToepfer_1985,
  author   = {A Toepfer and H J L\"udde and B Jacob and R M Dreizler},
  title    = {{Many-electron aspects in ion-atom collisions: 2p-2s vacancy transfer in the ${\mathrm{Ne}}^{+}\text{\ensuremath{+}}{\mathrm{Ne}}$ system}},
  journal  = {Journal of Physics B: Atomic and Molecular Physics},
  year     = {1985},
  volume   = {18},
  number   = {10},
  pages    = {1969},
  month    = {may},
  doi      = {10.1088/0022-3700/18/10/014},
  url      = {https://dx.doi.org/10.1088/0022-3700/18/10/014},
}

@Article{Kim2014,
  author    = {Kim, H.-K. and Gassert, H. and Titze, J. N. and Waitz, M. and Voigtsberger, J. and Trinter, F. and Becht, J. and Kalinin, A. and Neumann, N. and Zhou, C. and Schmidt, L. Ph. H. and Jagutzki, O. and Czasch, A. and Sch\"offler, M. and Merabet, H. and Schmidt-B\"ocking, H. and Jahnke, T. and L\"udde, H. J. and Cassimi, A. and D\"orner, R.},
  title     = {{Orientation dependence in multiple ionization of ${\mathrm{He}}_{2}$ and ${\mathrm{Ne}}_{2}$ induced by fast, highly charged ions: Probing the impact-parameter-dependent ionization probability in 11.37-MeV/u ${S}^{14+}$ collisions with He and Ne}},
  journal   = {Phys. Rev. A},
  year      = {2014},
  volume    = {89},
  pages     = {022704},
  month     = {Feb},
  issue     = {2},
  numpages  = {11},
  publisher = {American Physical Society},
  url       = {https://link.aps.org/doi/10.1103/PhysRevA.89.022704},
}

@Article{Iskandar2014,
  author    = {Iskandar, W. and Matsumoto, J. and Leredde, A. and Fl\'echard, X. and Gervais, B. and Guillous, S. and Hennecart, D. and M\'ery, A. and Rangama, J. and Zhou, C. L. and Shiromaru, H. and Cassimi, A.},
  title     = {{Atomic Site-Sensitive Processes in Low Energy Ion-Dimer Collisions}},
  journal   = {Phys. Rev. Lett.},
  year      = {2014},
  volume    = {113},
  pages     = {143201},
  month     = {Oct},
  doi       = {10.1103/PhysRevLett.113.143201},
  issue     = {14},
  numpages  = {5},
  publisher = {American Physical Society},
  url       = {https://link.aps.org/doi/10.1103/PhysRevLett.113.143201},
}

@Article{Ludde_1985,
  author   = {H J L\"udde and R M Dreizler},
  title    = {Comment on inclusive cross sections},
  journal  = {Journal of Physics B: Atomic and Molecular Physics},
  year     = {1985},
  volume   = {18},
  number   = {1},
  pages    = {107},
  month    = {jan},
  doi      = {10.1088/0022-3700/18/1/012},
  url      = {https://dx.doi.org/10.1088/0022-3700/18/1/012},
}

@Article{Starko_2025,
  author    = {Starko, Darij and Kirchner, Tom},
  title     = {{Investigations of electron removal processes in slow ${\mathrm{He}}^{2+}$- and ${\mathrm{He}}^{+}$-${\mathrm{Ne}}_{2}$ collisions and of their implications for the subsequent dimer fragmentation through interatomic Coulombic decay}},
  journal   = {Journal of Physics B: Atomic, Molecular and Optical Physics},
  year      = {2025},
  volume    = {58},
  number    = {7},
  pages     = {075204},
  month     = {Apr},
  doi       = {10.1088/1361-6455/adc522},
  publisher = {IOP Publishing},
  url       = {https://dx.doi.org/10.1088/1361-6455/adc522},
}

@Article{Zhang_2025,
  author    = {Zhang, Yu and Wang, Jiarong and Hu, Xiaoqing and Ren, Baihui and Meng, Tianming and Ma, Pufang and Qi, Yueying and Wu, Yong and Wang, Jianguo and Zou, Yaming and Tu, Bingsheng and Cassimi, Amine and Wei, Baoren},
  title     = {Nonadiabatic-coupling-mediated argon dimer dissociation by slow and low-charge-state ion collisions},
  journal   = {Phys. Rev. A},
  year      = {2025},
  volume    = {111},
  pages     = {042824},
  month     = {Apr},
  doi       = {10.1103/PhysRevA.111.042824},
  issue     = {4},
  numpages  = {5},
  publisher = {American Physical Society},
}

@Article{OJKroneisen_1999,
  author   = {O J Kroneisen and H J Lüdde and T Kirchner and R M Dreizler},
  title    = {The basis generator method: optimized dynamical representation of the solution of time-dependent quantum problems},
  journal  = {Journal of Physics A: Mathematical and General},
  year     = {1999},
  volume   = {32},
  number   = {11},
  pages    = {2141},
  month    = {Mar},
  doi      = {10.1088/0305-4470/32/11/009},
}

@Misc{NIST_ASD,
  author       = {A.~Kramida and {Yu.~Ralchenko} and J.~Reader and {and NIST ASD Team}},
  howpublished = {{NIST Atomic Spectra Database (ver. 5.12), [Online]. Available: {\tt{https://physics.nist.gov/asd}} [2025, December 4]. National Institute of Standards and Technology, Gaithersburg, MD.}},
  year         = {2024},
}

@Article{Alizadeh_2015,
  author  = {Alizadeh, E. and Orlando, T. and Sanche, L.},
  title   = {Biomolecular Damage Induced by Ionizing Radiation: The Direct and Indirect Effects of Low-Energy Electrons on DNA},
  journal = {Annual Review Physical Chemistry},
  year    = {2015},
  volume  = {66},
  pages   = {379-398},
}

@Article{Vendrell_2010,
  author   = {Vendrell, Oriol and Stoychev, Spas D. and Cederbaum, Lorenz S.},
  title    = {Generation of Highly Damaging H2O+ Radicals by Inner Valence Shell Ionization of Water},
  journal  = {ChemPhysChem},
  year     = {2010},
  volume   = {11},
  number   = {5},
  pages    = {1006-1009},
  doi      = {https://doi.org/10.1002/cphc.201000034},
  eprint   = {https://chemistry-europe.onlinelibrary.wiley.com/doi/pdf/10.1002/cphc.201000034},
  url      = {https://chemistry-europe.onlinelibrary.wiley.com/doi/abs/10.1002/cphc.201000034},
}

@Article{Pettersson_2013,
  author  = {Pettersson, L},
  title   = {Radical water},
  journal = {Nature Chem},
  year    = {2013},
  volume  = {5},
  pages   = {553-554},
}

@Article{Moseley_1977,
  author  = {Moseley, J. T. and Saxon, R. P. and Huber, B. A. and Cosby, P. C. and Abouaf, R. and Tadjeddine, M.},
  title   = {{Photofragment spectroscopy and potential curves of $\mathrm{Ar}^{+}_{2}$}},
  journal = {The Journal of Chemical Physics},
  year    = {1977},
  volume  = {67},
  number  = {4},
  pages   = {1659-1668},
  month   = {08},
  doi     = {10.1063/1.434998},
  issn    = {0021-9606},
  url     = {https://doi.org/10.1063/1.434998},
}

@Article{Kimura_2013,
  author    = {Kimura, M. and Fukuzawa, H. and Sakai, K. and Mondal, S. and Kukk, E. and Kono, Y. and Nagaoka, S. and Tamenori, Y. and Saito, N. and Ueda, K.},
  title     = {Efficient site-specific low-energy electron production via interatomic Coulombic decay following resonant Auger decay in argon dimers},
  journal   = {Phys. Rev. A},
  year      = {2013},
  volume    = {87},
  pages     = {043414},
  month     = {Apr},
  doi       = {10.1103/PhysRevA.87.043414},
  issue     = {4},
  numpages  = {4},
  publisher = {American Physical Society},
  url       = {https://link.aps.org/doi/10.1103/PhysRevA.87.043414},
}

\end{document}